\documentclass[%
 reprint,amsmath,amssymb, aps,]{revtex4-1}
\usepackage{rotating,graphicx}% Include figure files
\usepackage{dcolumn}% Align table columns on decimal point
\usepackage{bm}% bold math
\usepackage{amsmath}
\usepackage{amssymb}
\usepackage{xcolor}

\begin{document}
\title{Topological photonic crystal fibers and ring resonators}

\author{Laura Pilozzi}%
\email{laura.pilozzi@isc.cnr.it}
\affiliation{Institute for Complex Systems, National Research Council (ISC-CNR), Via dei Taurini 19, 00185 Rome, Italy
}%

\author{Daniel Leykam}%
\affiliation{
Center for Theoretical Physics of Complex Systems, Institute for Basic Science (IBS), Daejeon 34126, Republic of Korea}

\author{Zhigang Chen}%
\affiliation{
The Key Laboratory of Weak-Light Nonlinear Photonics, TEDA Applied Physics
Institute, Nankai University, Tianjin 300457, China}
\affiliation{
	Department of Physics and Astronomy, San Francisco State University, San Francisco, California 94132, USA}

\author{Claudio Conti}
\affiliation{
 Institute for Complex Systems, National Research Council (ISC-CNR),  Via dei Taurini 19, 00185 Rome, Italy
}%
\affiliation{
  Department of Physics, University Sapienza, Piazzale Aldo Moro 5, 00185 Rome, Italy
}%
\affiliation{
The Key Laboratory of Weak-Light Nonlinear Photonics, TEDA Applied Physics
Institute, Nankai University, Tianjin 300457, China}

\graphicspath{{Images/}}

\begin{abstract}
We study photonic crystal fibers and ring resonators with topological features
induced by Aubry-Andre-Harper modulations of the cladding. We find non trivial gaps and edge states at the interface between regions with different Chern numbers.  We calculate the field profile and eigenvalue dispersion by an exact recursive approach. 
Compared with conventional circular resonators and fibers, the proposed structure features topological protection and hence robustness against symmetry-preserving local perturbations that do not close the gap.
These topological photonic crystal fibers sustain strong field localization and energy concentration at a given radial distance. As topological light guiding and trapping devices, they may bring about many opportunities for both fundamentals and applications unachievable with conventional optical devices.
\end{abstract}


\maketitle

%\section{\label{sec:level1}Introduction}
The seminal papers on analogs of quantum Hall effect in optics~\cite{Raghu,Haldane} boosted the research on photonic systems described by magnetic-like Hamiltonians~\cite{Wang,Wang2,Hafezi1,Fang,Skirlo,Hafezi2,Khanikaev,Rechtsman,Longhi,Zeng}.
Photonic topological insulators hold great promises for applications, such as robust unidirectional propagation of light and topological  lasers~\cite{Pilozzi,Amo,Hararil,Bahari}, topologically protected frequency combs \cite{Pilozzi2017}, topological nanostructures\cite{Kruk2019, Koshelev2019}, and quantum information processing~\cite{Mittal,Rechtsman2,Mittal3}. Recent developments include the applications of machine learning to design topological devices \cite{Pilozzi2018, Long2019}. Edge states with energies lying in the bulk gaps are the leading ingredient in these applications. Edge states are localized at the boundary between regions with diverse topological invariants~\cite{Thouless} as the first Chern number of the bands - or the winding number of the gaps~\cite{Hatsugai} - and are robust against backscattering from impurities.

Nontrivial topological phases also arise in a photonic system by employing synthetic dimensions~\cite{Ozawa,Yuan,Lustig}. An example is the Aubry-Andre-Harper (AAH) modulation~\cite{Harper,Aubry,Kraus,Ganeshan} of optical lattice parameters. One-dimensional (1D) systems with synthetic dimensions have the same topological features of their 2D periodic ancestor lattices~\cite{Hofstadter,Posha}. However, the application of synthetic dimensions has been so far limited to linear geometries. The use of AAH and similar strategies for synthetic dimensions in circular, elliptical, or more complex coordinates is unexplored.

In this Letter, we introduce the concept of topological photonic crystal fibers and resonators that exploit topological features due to the AAH modulation in cylindrical symmetry to guide and trap electromagnetic radiation on edge states.
These new optical devices support tightly confined modes in the radial directions, protected with respect to disorder and other perturbations, such as bending.

Conventional circular optical waveguides and resonators exploit total internal reflection (TIR) with
a solid core surrounded by a lower refractive index medium, or Bragg or photonic crystal claddings~\cite{Yariv,Russel,Cregan,Knight}.
Photonic crystal fibers (PCFs) guide light in a hollow-core by Bragg reflection; as a result, bending losses are strongly reduced for a large core.
Reconfigurable index structures, similar to Bragg fibers, can also be obtained
by Bessel photonic lattices in bulk crystals~\cite{Wang2006}. One-way fiber modes, protected against backscattering, in a 3D magnetic Weyl photonic crystal, can be realized at microwave frequencies~\cite{Lu2018}.
Twisting and deforming
PCFs enable the control of the angular momentum~\cite{Wong2012} and induce optomechanical nonlinearities~\cite{butsch2012}. Hollow-core PCFs may guide dielectric particles by radiation pressure~\cite{garbos2011}.
Also, the photonic bandgap mechanism allows annular Bragg resonators~\cite{Scheuer}.

Our proposal of a cylindrical photonic topological insulator, sustaining edge states, is a recipe for the cladding of PCFs to gain topological protection.
In our configuration, the core-cladding interface acts as the boundary between two distinct topological phases: a trivial one, the core, and a topological one, the cladding with radiative edge states~\cite{Posha}.
This boundary can also be attained inside the cladding if a radial distance $\rho_n$ value divides the cladding in two substructures with different modulations.

The structure considered has a core of dielectric constant $\epsilon_c$ with a radius $\rho_1$ and a cladding given by a sequence of two homogeneous layers A and B characterized by dielectric functions $\epsilon_a$ and $\epsilon_b$. The center positions of the A layers, $s_a$ wide, are given by $\rho_n^A = d_o\left[ {n + \eta \delta_n^H } \right]$, where $\delta_n^H = \cos (2\pi\gamma n + \phi )$ is the Harper modulation~\cite{Harper} with $\gamma=p/q$ and p and q coprime integer numbers~\cite{Hofstadter}; $\eta$ is a coefficient that controls the modulation strength. 
The cladding is then, in the $\rho$ direction, a periodic structure with q layers A in the unit cell and a period $d = q d_o$, where $d_o$ is the period of the unmodulated structure ($ \eta = 0$). The phase $\phi$, the topological parameter of the 1D periodic modulation, varying in (0, $2\pi$), adiabatically deforms the system and accounts for the momentum along the second geometrical dimension of the 2D ancestor lattice~\cite{Hofstadter,Posha}. 
\\
For this sequence of annular regions, the interfaces positions can be written as $\rho_j=\rho_j'+\delta$ where:
\[\rho_j'=\rho^A_{ \lfloor\frac{j+1}{2} \rfloor}+(-1)^{j-2 \lfloor\frac{j-1}{2}\rfloor}s_a/2\]
$\delta=\rho_1-\rho_1'$ and $\lfloor x \rfloor$ is the integer part of x.

Different methods for the analysis of structured claddings have been proposed in the literature, ranging from approximated ones like the method using asymptotic approximations of Bessel functions~\cite{Xu,Xu2}, to exact ones like the standard transfer matrix method~\cite{Yariv}. 
We choose to apply the exact recursive approach proposed by Chew~\cite{Chew} to analize the bandgap structure and design the resonator, since, casting the problem in a 2x2 matrix form for the longitudinal components of the electric and magnetic fields, allows to obtain, in a straightforward way, the complex mode frequencies.
Moreover, this recursive formalism enables us to find the modal distribution in the case of an arbitrary arrangement of annular concentric regions and so to study the effects of shallow disorder and verify the topological protection.
 
\begin{figure}[t]
		\includegraphics[width=1\columnwidth]{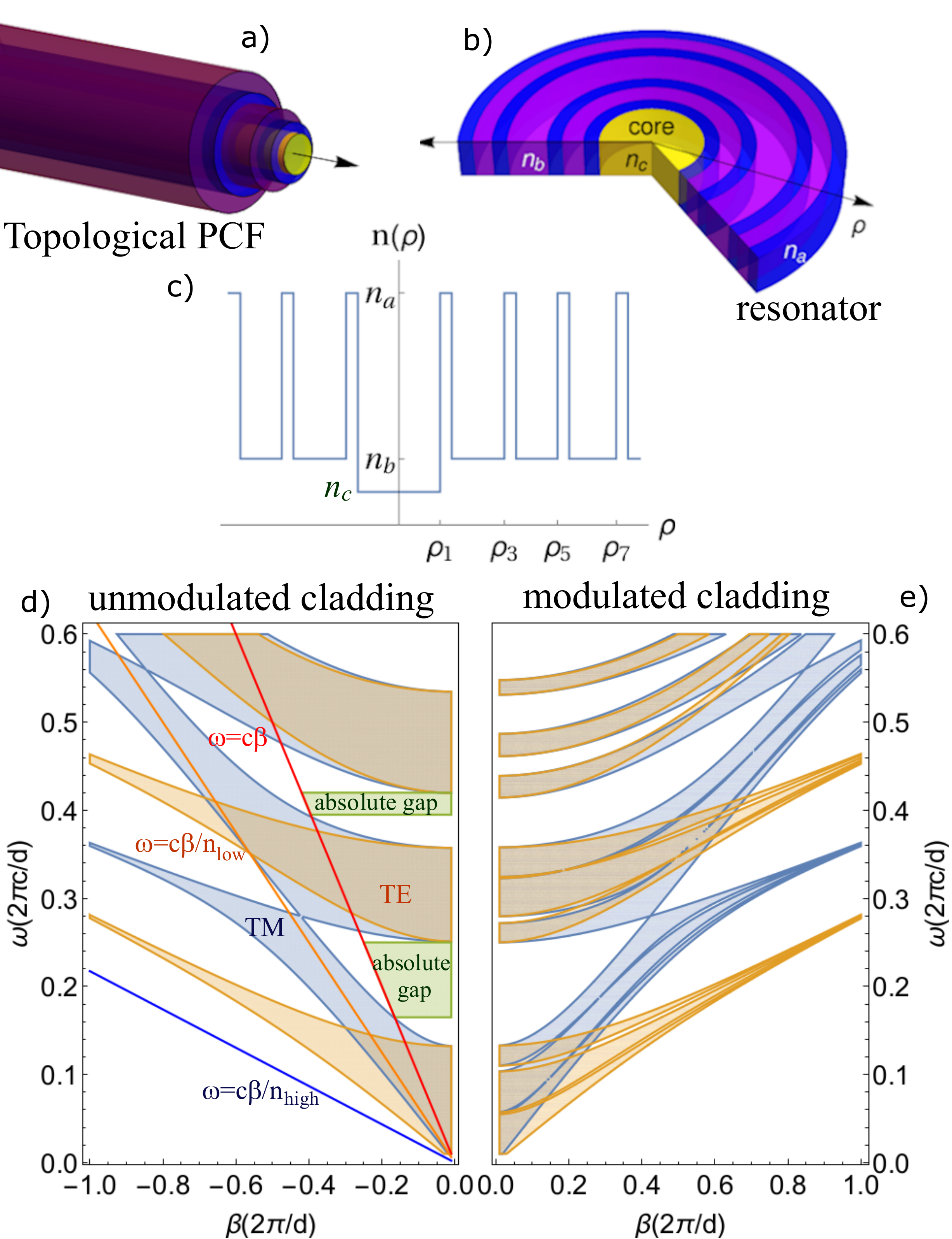}	
	\caption{(Color online) Schematic representation of a) the topological PCF and b) the optical resonator. c) Piecewise dielectric function profile for an AAH cladding with $\gamma=1/3$, taking on the values $n_c$ for the core, and $n_a$, $n_b$ for the cladding. Band structure in the asymptotic limit for d) an unmodulated and e) a modulated cladding with $n_a=n_{high}$=4.6 (tellurium), $n_b=n_{low}=1.6$  (polystyrene), $s_a=0.33d_o$, $s_b=0.67d_o$; $n_c$=1. Filled regions represent the modes that propagate within the multilayer cladding for TE (orange) and TM (blue) polarization. \label{fig:f1}}
\end{figure}

We consider cylindrical coordinates and waves traveling in the z-direction with propagation constant $\beta$, so every field component has the form: $\psi (\rho ,\vartheta ,z,t) = \psi (\rho ,\vartheta )e^{i(\beta z - \omega t)}$  
where $\omega$ is the angular frequency.
\\
Figure~\ref{fig:f1}) schematically illustrates a) the topological PCF, b) the topological optical resonator and c) their radial index profile. In these systems the permittivity depends only on the radial coordinate: $\epsilon(\rho)=\epsilon_j$ for $\rho_{j-1}<\rho<\rho_j$ with $\rho_0=0$.
In each homogeneous cylindrical layer, specified by its dielectric function $\epsilon_j$, the longitudinal field components $E_z^j (\rho ,\vartheta )$ and $H_z^j (\rho ,\vartheta )$ are the solutions of the equation: 
\begin{equation}
\left[ {\frac{{\partial ^2 }}{{\partial \rho ^2 }} + \frac{1}{\rho }\frac{\partial }{{\partial \rho }} + (k_j^2  - \frac{{\ell ^2 }}{{\rho ^2 }})} \right]\left\{ {\begin{array}{*{20}c}
	{E_z^j (\rho ,\vartheta )}  \\
	{H_z^j (\rho ,\vartheta )}  \\
	\end{array}} \right\} = 0  \label{equ}
\end{equation}
where $l$ is an integer representing the angular modal number and $k_j^2=\frac{\omega^2}{c^2}\epsilon_j-\beta^2$, with c being the speed of light. 
We choose, as independent solutions of equation~\ref{equ}), the Bessel $J_\ell  (k_j \rho )$
and the Hankel $H_\ell ^{(1)} (k_j \rho )$ functions of the first kind so that inside each homogeneous cylindrical layer j the field can be written as the superposition of an outgoing wave $H_\ell ^{(1)} (k_j \rho$) and a standing wave $J_\ell  (k_j \rho )$~\cite{Chew}:
\begin{equation}
\left[ {\begin{array}{*{20}c}
	{E_z^j }  \\
	{H_z^j }  \\
	\end{array}} \right] = \left[ {H_\ell ^{(1)} (k_j \rho )\mathord{\buildrel{\lower3pt\hbox{$\scriptscriptstyle\leftrightarrow$}} 
		\over I}  + J_\ell  (k_j \rho )\tilde {\mathord{\buildrel{\lower3pt\hbox{$\scriptscriptstyle\leftrightarrow$}} 
		\over R}} _{j,j + 1} } \right]\vec a_j \label{eq:e2}
\end{equation}
Here $\vec a_j  \equiv \left[ {\begin{array}{*{20}c}
	{e_{jz} }  \\
	{h_{jz} }  \\
	\end{array}} \right]$ determines the relative amplitudes of the electric and magnetic field components propagating outwards and $
\tilde {\mathord{\buildrel{\lower3pt\hbox{$\scriptscriptstyle\leftrightarrow$}} 
		\over R}} _{j,j + 1} $, giving the relation between the electric and magnetic field components of the inward and outward propagating fields, is a generalized reflection matrix with $\tilde {\mathord{\buildrel{\lower3pt\hbox{$\scriptscriptstyle\leftrightarrow$}} 
		\over R}} _{N,N + 1}=0 $, where N is the number of layers in the structure.
\\
In the outermost layer N, only the outgoing wave is present:
\begin{equation}
\left[ {\begin{array}{*{20}c}
	{E_z^N }  \\
	{H_z^N }  \\
	\end{array}} \right] = H_\ell ^{(1)} (k_N \rho )\mathord{\buildrel{\lower3pt\hbox{$\scriptscriptstyle\leftrightarrow$}} 
	\over T} _{N-1,N} \vec a_{N-1} \label{eq:e3}
\end{equation}
while in the innermost layer, since $\lim_{x\to0} H_\ell ^{(1)} (x)=\infty$, we can only have the standing wave:
\begin{equation}
\left[ {\begin{array}{*{20}c}
	{E_z^1 }  \\
	{H_z^1 }  \\
	\end{array}} \right] =  J_\ell  (k_1 \rho )\tilde{\mathord{\buildrel{\lower3pt\hbox{$\scriptscriptstyle\leftrightarrow$}} 
		\over R}} _{1,2} \vec a_1 
\end{equation}

We observe that an outgoing wave in region j+1 is a consequence of the transmission of an outgoing wave in region j plus the reflection of a standing wave in region j+1:
\[
\vec a_{j + 1}  = \mathord{\buildrel{\lower3pt\hbox{$\scriptscriptstyle\leftrightarrow$}} 
	\over T} _{j,j + 1} \vec a_j  + \mathord{\buildrel{\lower3pt\hbox{$\scriptscriptstyle\leftrightarrow$}} 
	\over R} _{j + 1,j} \tilde{\mathord{\buildrel{\lower3pt\hbox{$\scriptscriptstyle\leftrightarrow$}} 
		\over R}} _{j + 1,j + 2} \vec a_{j + 1} 
\]
\\
Moreover a standing wave in region j is the result of an outgoing wave in region j plus a standing wave in region j+1:
\[
\tilde{ \mathord{\buildrel{\lower3pt\hbox{$\scriptscriptstyle\leftrightarrow$}} 
		\over R}} _{j,j + 1} \vec a_j  = \mathord{\buildrel{\lower3pt\hbox{$\scriptscriptstyle\leftrightarrow$}} 
	\over R} _{j,j + 1} \vec a_j  + \mathord{\buildrel{\lower3pt\hbox{$\scriptscriptstyle\leftrightarrow$}} 
	\over T} _{j + 1,j} \tilde {\mathord{\buildrel{\lower3pt\hbox{$\scriptscriptstyle\leftrightarrow$}} 
		\over R}} _{j + 1,j + 2} \vec a_{j + 1} 
\]
\\
As a consequence recursive relations can be obtained for the reflection matrices:
\begin{multline}
\tilde {\mathord{\buildrel{\lower3pt\hbox{$\scriptscriptstyle\leftrightarrow$}} 
		\over R}} _{j,j + 1}  = 
\mathord{\buildrel{\lower3pt\hbox{$\scriptscriptstyle\leftrightarrow$}} 
	\over R} _{j,j + 1}  + \\
+\mathord{\buildrel{\lower3pt\hbox{$\scriptscriptstyle\leftrightarrow$}} 
	\over T} _{j + 1,j} \tilde{ \mathord{\buildrel{\lower3pt\hbox{$\scriptscriptstyle\leftrightarrow$}} 
		\over R}} _{j + 1,j + 2} \left( {\mathord{\buildrel{\lower3pt\hbox{$\scriptscriptstyle\leftrightarrow$}} 
		\over I}  - \mathord{\buildrel{\lower3pt\hbox{$\scriptscriptstyle\leftrightarrow$}} 
		\over R} _{j + 1,j} \tilde{ \mathord{\buildrel{\lower3pt\hbox{$\scriptscriptstyle\leftrightarrow$}} 
			\over R}} _{j + 1,j + 2} } \right)^{ - 1} \mathord{\buildrel{\lower3pt\hbox{$\scriptscriptstyle\leftrightarrow$}} 
	\over T} _{j,j + 1} \label{ricor}
\end{multline}
and the field amplitudes:
\begin{equation}
\vec a_{j + 1}  = (\mathord{\buildrel{\lower3pt\hbox{$\scriptscriptstyle\leftrightarrow$}} 
	\over I}-\mathord{\buildrel{\lower3pt\hbox{$\scriptscriptstyle\leftrightarrow$}} 
	\over R} _{j + 1,j} \tilde{\mathord{\buildrel{\lower3pt\hbox{$\scriptscriptstyle\leftrightarrow$}} 
		\over R}} _{j + 1,j + 2}  )^{-1}\mathord{\buildrel{\lower3pt\hbox{$\scriptscriptstyle\leftrightarrow$}} 
	\over T} _{j,j + 1} \vec a_j  \label{ricor2} 
\end{equation}
for a N-layer cylinder~\cite{Chew}.
The generalized reflection matrix, eq.~\ref{ricor}), includes the effects of reflections and transmissions at all the layers beyond the j-th one. 
With this equation and starting from the outermost, one can find the reflection and transmission matrices for all the N layers; then, starting from the innermost layer, the field can be recursively described, through eq.~\ref{ricor2}), in the whole structure.
For the first interface, between core and cladding, the boundary conditions read:
\begin{equation}
\begin{array}{l}
J_\ell  (k_1 \rho _1 )\tilde{\mathord{\buildrel{\lower3pt\hbox{$\scriptscriptstyle\leftrightarrow$}} 
	\over R}} _{1,2}\vec{a}_1 =\left[H_\ell ^{(1)} (k_2 \rho _1 )\mathord{\buildrel{\lower3pt\hbox{$\scriptscriptstyle\leftrightarrow$}} 
	\over I}  + J_\ell  (k_2 \rho _1 )\tilde{\mathord{\buildrel{\lower3pt\hbox{$\scriptscriptstyle\leftrightarrow$}} 
	\over R}} _{2,3}\right]\vec{a}_2  \\ 
\mathord{\buildrel{\lower3pt\hbox{$\scriptscriptstyle\leftrightarrow$}} 
	\over J}_\ell  (k_1 \rho _1 )\tilde{\mathord{\buildrel{\lower3pt\hbox{$\scriptscriptstyle\leftrightarrow$}} 
		\over R}} _{1,2}\vec{a}_1 =\left[\mathord{\buildrel{\lower3pt\hbox{$\scriptscriptstyle\leftrightarrow$}} 
	\over H}_\ell ^{(1)} (k_2 \rho _1 )\mathord{\buildrel{\lower3pt\hbox{$\scriptscriptstyle\leftrightarrow$}} 
	\over I}  + \mathord{\buildrel{\lower3pt\hbox{$\scriptscriptstyle\leftrightarrow$}} 
	\over J}_\ell  (k_2 \rho _1 )\tilde{\mathord{\buildrel{\lower3pt\hbox{$\scriptscriptstyle\leftrightarrow$}} 
	\over R}} _{2,3}\right]\vec{a}_2 
\end{array}  
\end{equation}
where $ \vec a_2  = (\mathord{\buildrel{\lower3pt\hbox{$\scriptscriptstyle\leftrightarrow$}} 
	\over I}-\mathord{\buildrel{\lower3pt\hbox{$\scriptscriptstyle\leftrightarrow$}} 
	\over R} _{2,1} \tilde{\mathord{\buildrel{\lower3pt\hbox{$\scriptscriptstyle\leftrightarrow$}} 
		\over R}} _{2,3}  )^{-1}\mathord{\buildrel{\lower3pt\hbox{$\scriptscriptstyle\leftrightarrow$}} 
	\over T} _{1,2} \vec a_1  $
and:

\begin{multline}
\tilde {\mathord{\buildrel{\lower3pt\hbox{$\scriptscriptstyle\leftrightarrow$}} 
		\over R}} _{1,2}  =\frac{1}{J_\ell  (k_1 \rho _1 )}\\
\left[H_\ell ^{(1)} (k_2 \rho _1 )\mathord{\buildrel{\lower3pt\hbox{$\scriptscriptstyle\leftrightarrow$}} 
	\over I}  + J_\ell  (k_2 \rho _1 )\tilde{\mathord{\buildrel{\lower3pt\hbox{$\scriptscriptstyle\leftrightarrow$}} 
	\over R}} _{2,3}\right] 
\left( {\mathord{\buildrel{\lower3pt\hbox{$\scriptscriptstyle\leftrightarrow$}} 
		\over I}  - \mathord{\buildrel{\lower3pt\hbox{$\scriptscriptstyle\leftrightarrow$}} 
		\over R} _{2,1} \tilde{ \mathord{\buildrel{\lower3pt\hbox{$\scriptscriptstyle\leftrightarrow$}} 
			\over R}} _{2,3} } \right)^{ - 1} \mathord{\buildrel{\lower3pt\hbox{$\scriptscriptstyle\leftrightarrow$}} 
	\over T} _{1,2} \label{eq:r12}
\end{multline}
\\
The boundary conditions, described in the Appendix~\ref{sec:bound}, for the fields at $\rho  = \rho _j$ between two adjacent cylindrical layers "j" and "j+1", determine the reflection and transmission matrices $\mathord{\buildrel{\lower3pt\hbox{$\scriptscriptstyle\leftrightarrow$}} 
	\over R} _{j,j + 1} $ and $\mathord{\buildrel{\lower3pt\hbox{$\scriptscriptstyle\leftrightarrow$}} 
	\over T} _{j,j + 1}$. 
\\
Guided modes, defined as the nontrivial solutions that exist without the need for an external excitation, can be found requiring that the reflection matrix have an infinite determinant.
\\
The guidance condition, given for a two layer structure in Appendix~\ref{sec:edge}, for an N-layer fiber can be obtained from the generalized reflection matrix in eq.~\ref{eq:r12}), and reads:
\begin{equation}
det\left( {\mathord{\buildrel{\lower3pt\hbox{$\scriptscriptstyle\leftrightarrow$}} 
		\over I}  - \mathord{\buildrel{\lower3pt\hbox{$\scriptscriptstyle\leftrightarrow$}} 
		\over R} _{2,1} \tilde{ \mathord{\buildrel{\lower3pt\hbox{$\scriptscriptstyle\leftrightarrow$}} 
			\over R}} _{2,3} } \right)\equiv f(\omega,\beta)=0\label{eq:condN}
\end{equation}
The solutions of eq.~(\ref{eq:condN}), giving rise to a set of complex $(\omega,\beta)$ states with either real frequency or real wave vector, are the allowed modes of the fiber.
\\
Unlike Bragg guiding fibers, the core-cladding interface acts as the boundary between two distinct topological phases: a trivial one, the core, and a topological one, the cladding with radiative edge states~\cite{Posha}, here described in the complex $\omega$ representation~\cite{Posha2} where the real part is the modal resonance frequency while the imaginary part is the decay rate accounting for the modal loss~\cite{Xu1}.
This description is equivalent~\cite{Huang} to the one with real frequency and complex propagation constant $\beta$, with the imaginary part of $\beta$ providing the radiative decay of the leaky modes~\cite{Yariv, Johnson}.
\begin{figure} [t]
	\includegraphics[width=1.\columnwidth]{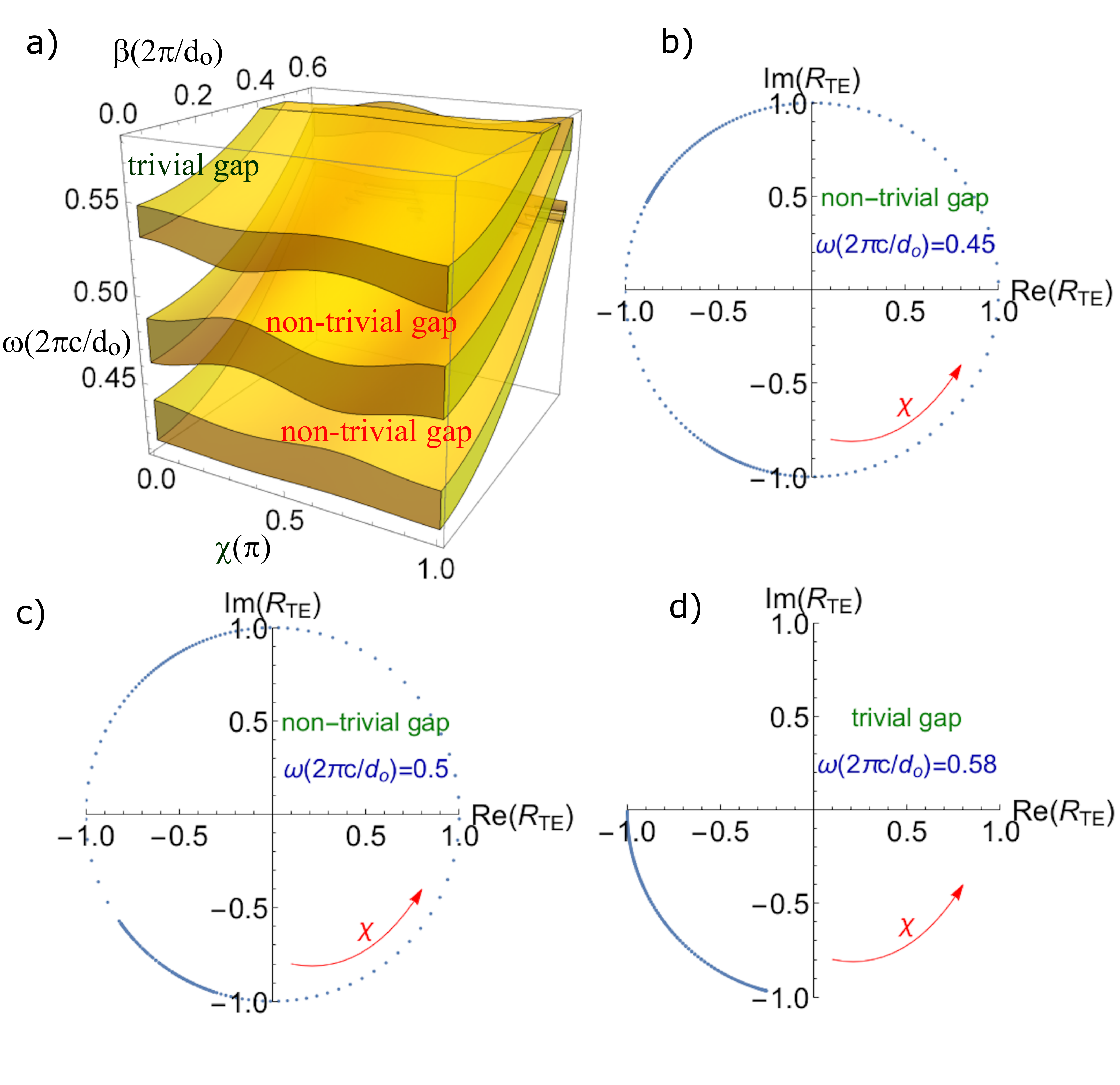}
	\caption{a) Band structure, for TE polarization, in the asymptotic limit for a modulated cladding as a function of $\chi$ and $\beta$. \\
		b), c), d) Winding numbers of the reflection coefficient for the three gaps.}\label{fig:bands}
\end{figure}
\begin{figure*} [t]
	\includegraphics[width=2\columnwidth]{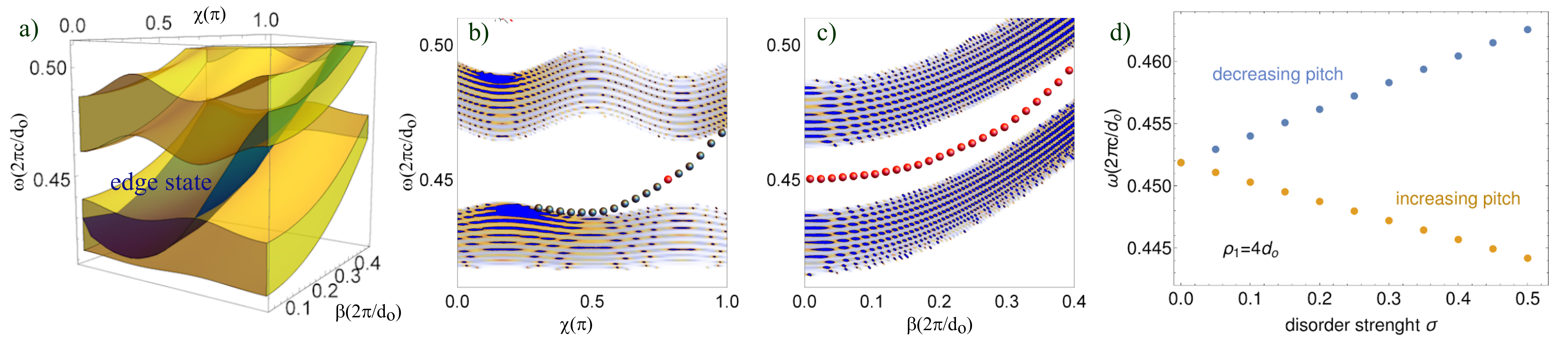}
	\caption{Reflectivity map and edge states for TE polarization a) in the asymptotic limit and the exact solution for b) $\beta=0.1 (2 \pi/d_o)$ and c) $\chi=0.8 \pi$.  In b) and c) the results are for a finite cladding with 13 unitary cells: the minimum number to make the gaps converging to the periodic case. d) Edge mode for $\beta=0.1 (2 \pi/d_o)$ and $\chi=0.8 \pi$ as a function of the disorder strength for two realizations of disorder: the orange(blu) curve is for a system with an increasing(decreasing) normalized period, d/$d_o$.}\label{fig:bands3}
\end{figure*}

Due to the periodic cladding structuring in the radial direction, our topological PCF shows a gapped spectrum. It can not be analyzed in the framework of the Bloch theorem due to the fact that, for radially depending dielectric functions, the operator $O=
\left[ {-\frac{1}{\rho \epsilon(\rho)}\frac{{\partial }}{{\partial \rho  }}\left(\frac{\rho}{ \mu(\rho)}\frac{{\partial }}{{\partial \rho  }}\right) + \frac{\ell}{\rho^2 \epsilon(\rho) \mu(\rho)}} \right]$ is not invariant under translations. However, as detailed in Appendix~\ref{sec:asym}, one can always identify, in the cladding, a $\rho_n$ value such that for $\rho>\rho_n$, the structure shows gaps converging to that of a planar structure with the same Harper modulation.
\\
For a cladding consisting of alternating layers of high ($n_a=n_{high}$) and low ($n_b=n_{low}$) refractive indeces, Fig.\ref{fig:f1}, d) and e), show the one-dimensional gaps and bands, in the asymptotic limit, obtained through the transfer matrices (eq.s~\ref{eq:tjjs1} and \ref{eq:tjjs2}): d) shows the case of a periodic unmodulated cladding ($\eta=0$) while e) includes the Harper modulation. This modulation adds additional gaps where we expect to find guided modes as well as the edge states; the gap dependence on the shifted phase $\chi=\phi+\frac{\pi}{2}(2\gamma-1)$ is shown in Fig.~\ref{fig:bands}a) for the case of TE polarization, in the adimensional energy range (0.41,0.58).
\\
Both the plots in Fig.~\ref{fig:f1} and Fig.~\ref{fig:bands} are obtained through the quantity $2\xi =-Tr[T(\omega,\beta,\chi)]$, involving the trace of the transfer matrix $T(\omega,\beta,\chi)$ in the asymptotic limit, allowing one to locate the cladding gaps in the regions where $\xi^2 >2$.
\\
To take into account the finite value of the core radius $\rho_1$, affecting the actual position of the allowed modes, gaps ($|\tilde{\textbf R}_{1,2}|^2=1$) and bands ($|\tilde{\textbf R}_{1,2}|^2<1$) of the system can be equally well located through the map of the reflectivity modulus $|\tilde{\textbf R}_{1,2}(\omega,\beta, \chi)|^2$. The generalized reflectivity, $\tilde{\textbf R}_{1,2}(\omega,\beta, \chi)$, at the core-cladding interface also allows to define the nature of the gaps~\cite{Posha2}.

In Figs.~\ref{fig:bands} b), c) and d) we show, for the three different gaps of Fig.~\ref{fig:bands} a), the topological invariants given as the winding numbers $w_i$ of the reflection coefficient, i.e. the extra phase acquired by the reflectivity when $\chi$ varies in the range $(-\pi,\pi)$  while $\omega$ remains inside the gap. A nonzero winding number corresponds to a topologically nontrivial sample and is tied to the existence of topological edge states. So, as indicated in Fig.~\ref{fig:bands}a), unlike the upper one, the two lower gaps are nontrivial.
Moreover, through the reflectivity poles, edge state dispersions can be calculated. The one in the lower gap is shown in Fig.~\ref{fig:bands3}a) as a blue curve, bridging the bandgap in a given $(\chi, \beta)$ range.

Figs.~\ref{fig:bands3}b) and c) show the reflectivity map for TE polarization in the plane ($Re(\omega),\chi$) (b)) and in the plane ($Re(\omega),\beta$) (c)) for a structure with the same parameters of Fig.\ref{fig:bands} but a finite core radius $\rho_1=2d_o$. In this figure topological edge states are clearly seen: their real part $Re(\omega)$ i shown as a dotted curve in the gap while $Im(\omega)\approx 10^{-2}Re(\omega)$.

Their hallmark property of immunity to backscattering owing to topological protection is verified by introducing disorder in our structure as a randomized perturbation of the A layers center positions. 
In this case $\rho^A_n=d_o(n+\eta(\delta_n^H+\sigma\xi_n))$, where $\xi_n$ are random variables chosen in the range (-1,1), while $\sigma$ is the disorder strength.
Fig.~\ref{fig:bands3}d) shows the frequency variation of a specific mode for two realizations of increasing random disorder; it proves that it is nearly unaffected due to the topological protection, even for large perturbations ($\sigma\simeq~0.5$).
The different behavior for the two realizations is mainly due to the lattice pitch variation with disorder strenght: for the orange curve the lattice pitch increases while for the blue one the pitch decreses. As a consequence the whole spectrum shift to higher(lower) energies for decreasing(increasing) pitch.
 	Curves for different core radius values show unnoticeable differences since the mode frequencies mainly depend on the cladding features.

\begin{figure} [t]
	\includegraphics[width=0.8\columnwidth]{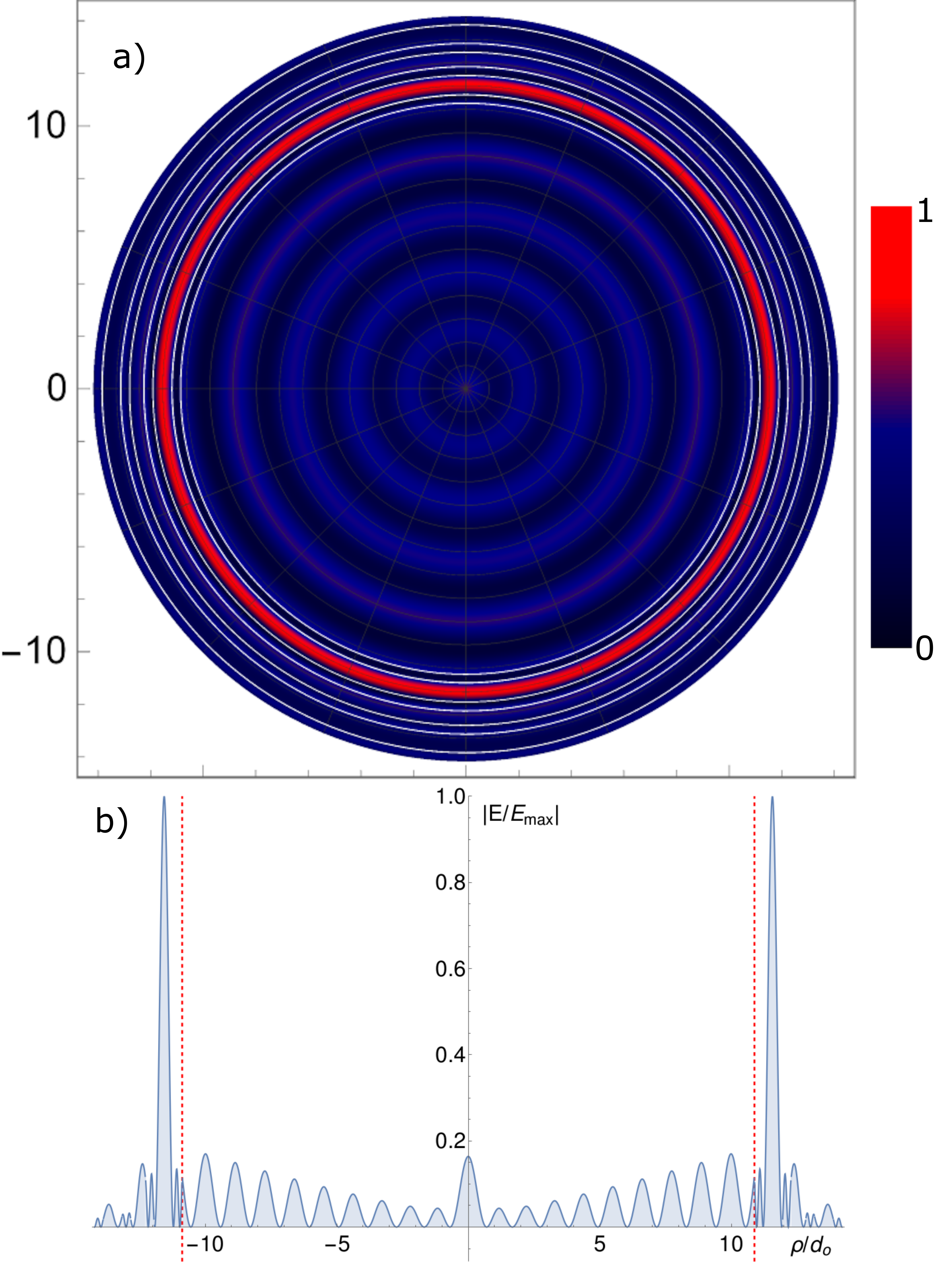} 	
	\caption{TE edge mode normalized field profile for the topological PCF described in Fig.~\ref{fig:f1} for $\chi=0.8 \pi$ and $\beta=0.1 (2 \pi/d_o)$. The dashed red line in b) marks the boundaries of the core region.}\label{fig:camp}
\end{figure}

Figure~\ref{fig:camp} displays, as an example, the normalized E-field pattern corresponding to the edge mode for $\chi=0.8 \pi$ and $\beta=0.1 (2 \pi/d_o)$, $\omega(2\pi c/d_o)=0.452+0.6*10^{-2}i$. This $l=0$ mode, unlike TIR or bandgap PCFs, is strongly localized at the boundary between core and cladding.

We theoretically studied topological band structures and edge-localized resonances in a cylindrical geometry with AAH modulation of the refractive index. A rigorous recursive approach - with no approximation - reveals edge modes with topological dispersion in the gapped spectrum. These edge states are robust to disorder and strongly localized in the radial direction. 
Optical fibers with topologically protected states open many new perspectives in the transmission of information for classical and quantum applications.
The resilience to external perturbations enables low-loss transport, quantum transport of non-classical states, and low-threshold topological-fiber-lasers.  Future directions also include the study of topological eigenmodes with angular momentum for multilevel signals, the study of the interplay between the Berry phase in twisted fibers and Chern numbers in synthetic dimensions, topological ring resonators for frequency comb generation, and linear and nonlinear metasurfaces by single and coupled resonators with radial Harper modulations.  Hollow-core fibers can accelerate dielectric and metallic particles; the way topological physics alters radiation-pressure and related phenomena - as optomechanical nonlinearities - is a new research direction. We also envisage many new physical phenomena when considering multimodal optical fibers with topologically protected states. 

We acknowledge support from the QuantERA ERA-NET Co-fund 731473 (Project QUOMPLEX), H2020
project grant number 820392, Sapienza Ateneo, PRIN 2015 NEMO (2015KEZNYM), PRIN 2017 PELM (20177PSCKT), Joint Bilateral Scientic Cooperation CNR-Italy/RFBR-Russia 2018-2020.
\\
ZC is supported by the National key R\&D Program of China under Grant (No. 2017YFA0303800).
\\
DL is supported by the Institute for Basic Science in Korea (IBS-R024-Y1).

\appendix

\section{\label{sec:bound}Boundary conditions}

The boundary conditions for the fields at $\rho  = \rho _j$ between two adjacent cylindrical layers "j" and "j+1", determine the reflection $\mathord{\buildrel{\lower3pt\hbox{$\scriptscriptstyle\leftrightarrow$}} 
	\over R} _{j,j + 1} $ and transmission $\mathord{\buildrel{\lower3pt\hbox{$\scriptscriptstyle\leftrightarrow$}} 
	\over T} _{j,j + 1}$ matrices. These conditions require the continuity of the longitudinal $(E_z^j ,H_z^j )$
components and of the transverse ones:
\begin{multline}
E_\vartheta ^j  = \frac{i}{{k_j^2 }}\left( {\omega \mu \frac{{\partial H_z^j }}{{\partial \rho }} - \frac{\beta }{\rho }\frac{{\partial E_z^j }}{{\partial \vartheta }}} \right)\\ H_\vartheta ^j  = \frac{i}{{k_j^2 }}\left( { - \omega \varepsilon \frac{{\partial E_z^j }}{{\partial \rho }} - \frac{\beta }{\rho }\frac{{\partial H_z^j }}{{\partial \vartheta }}} \right)
\end{multline}
that, trough eq.~\ref{eq:e2} and ~\ref{eq:e3}, can be written as:
\begin{equation}
\left[ {\begin{array}{*{20}c}
	{H_\vartheta ^j }  \\
	{E_\vartheta ^j }  \\
	\end{array}} \right] = \left[ {\mathord{\buildrel{\lower3pt\hbox{$\scriptscriptstyle\leftrightarrow$}} 
		\over H} _\ell ^{(1)} (k_j \rho ) + \mathord{\buildrel{\lower3pt\hbox{$\scriptscriptstyle\leftrightarrow$}} 
		\over J} _\ell  (k_j \rho )\mathord{\buildrel{\lower3pt\hbox{$\scriptscriptstyle\leftrightarrow$}} 
		\over R} _{j,j + 1} } \right]\vec a_j 
\end{equation}
\begin{equation}
\left[ {\begin{array}{*{20}c}
	{H_\vartheta ^{j + 1} }  \\
	{E_\vartheta ^{j + 1} }  \\
	\end{array}} \right] = \mathord{\buildrel{\lower3pt\hbox{$\scriptscriptstyle\leftrightarrow$}} 
	\over H} _\ell ^{(1)} (k_{j + 1} \rho )\mathord{\buildrel{\lower3pt\hbox{$\scriptscriptstyle\leftrightarrow$}} 
	\over T} _{j,j + 1} \vec a_j 
\end{equation}
where 
\begin{equation}
\mathord{\buildrel{\lower3pt\hbox{$\scriptscriptstyle\leftrightarrow$}} 
	\over B} _\ell  (k_j \rho ) =\frac{1}{{k_j^2 \rho }}\left[ {\begin{array}{*{20}c}
	{i\omega \varepsilon _j k_j \rho B_\ell'(k_j \rho )} & { - \ell \beta B_\ell  (k_j \rho ) }\\
	{ - \ell \beta B_\ell  (k_j \rho )} & { - i\omega \mu _j k_j \rho B_\ell' (k_j \rho )}  \\
	\end{array}} \right]\label{eq:eqb}
\end{equation}
with $B_\ell  (k_j \rho ) = H_\ell ^{(1)} (k_j \rho ),J_\ell  (k_j \rho )$.
\\
Moreover: $E_\rho ^j  = \frac{i}{{k_j^2 }}\left( { - \frac{{\omega \mu }}{\rho }\frac{{\partial H_z^j }}{{\partial \vartheta }} - \beta \frac{{\partial E_z^j }}{{\partial \rho }}} \right)$
and $H_\rho ^j  = \frac{i}{{k_j^2 }}\left( {\frac{{\omega \varepsilon }}{\rho }\frac{{\partial E_z^j }}{{\partial \vartheta }} - \beta \frac{{\partial H_z^j }}{{\partial \rho }}} \right)$.
\\
For the case of an outgoing wave incident from medium j on the boundary at $\rho_j$, from the continuity of $(E_z^j ,H_z^j )$, we have:
\begin{equation}
\left[ {H_\ell ^{(1)} (k_j \rho _j )\mathord{\buildrel{\lower3pt\hbox{$\scriptscriptstyle\leftrightarrow$}} 
		\over I}  + J_\ell  (k_j \rho _j )\mathord{\buildrel{\lower3pt\hbox{$\scriptscriptstyle\leftrightarrow$}} 
		\over R} _{j,j+1} } \right]\vec a_j  = H_\ell ^{(1)} (k_{j+1} \rho _j )\mathord{\buildrel{\lower3pt\hbox{$\scriptscriptstyle\leftrightarrow$}} 
	\over T} _{j,j+1} \vec a_j
\end{equation}
while, from the continuity of $(E_\vartheta ^j ,H_\vartheta ^j )$:
\[
\left[ {\mathord{\buildrel{\lower3pt\hbox{$\scriptscriptstyle\leftrightarrow$}} 
		\over H} _\ell ^{(1)} (k_j \rho _j ) + \mathord{\buildrel{\lower3pt\hbox{$\scriptscriptstyle\leftrightarrow$}} 
		\over J} _\ell  (k_j \rho _j )\mathord{\buildrel{\lower3pt\hbox{$\scriptscriptstyle\leftrightarrow$}} 
		\over R} _{j,j+1} } \right]\vec a_j  = \mathord{\buildrel{\lower3pt\hbox{$\scriptscriptstyle\leftrightarrow$}} 
	\over H} _\ell ^{(1)} (k_{j+1} \rho _j )\mathord{\buildrel{\lower3pt\hbox{$\scriptscriptstyle\leftrightarrow$}} 
	\over T} _{j,j+1} \vec a_j 
\]
Since $\vec a_j $ is nonzero we have:
\[
\begin{array}{l}
H_\ell ^{(1)} (k_j \rho _j )\mathord{\buildrel{\lower3pt\hbox{$\scriptscriptstyle\leftrightarrow$}} 
	\over I}  + J_\ell  (k_j \rho _j )\mathord{\buildrel{\lower3pt\hbox{$\scriptscriptstyle\leftrightarrow$}} 
	\over R} _{j,j+1}  = H_\ell ^{(1)} (k_{j+1} \rho _j )\mathord{\buildrel{\lower3pt\hbox{$\scriptscriptstyle\leftrightarrow$}} 
	\over T} _{j,j+1}  \\ 
\mathord{\buildrel{\lower3pt\hbox{$\scriptscriptstyle\leftrightarrow$}} 
	\over H} _\ell ^{(1)} (k_j \rho _j ) + \mathord{\buildrel{\lower3pt\hbox{$\scriptscriptstyle\leftrightarrow$}} 
	\over J} _\ell  (k_j \rho _j )\mathord{\buildrel{\lower3pt\hbox{$\scriptscriptstyle\leftrightarrow$}} 
	\over R} _{j,j+1}  = \mathord{\buildrel{\lower3pt\hbox{$\scriptscriptstyle\leftrightarrow$}} 
	\over H} _\ell ^{(1)} (k_{j+1} \rho _j )\mathord{\buildrel{\lower3pt\hbox{$\scriptscriptstyle\leftrightarrow$}} 
	\over T} _{j,j+1}  \\ 
\end{array}
\]
that can be solved to find  $\mathord{\buildrel{\lower3pt\hbox{$\scriptscriptstyle\leftrightarrow$}} 
	\over R} _{j,j+1}$ and $\mathord{\buildrel{\lower3pt\hbox{$\scriptscriptstyle\leftrightarrow$}} 
	\over T} _{j,j+1} $:

\begin{multline}
\left[ {H_\ell ^{(1)} (k_{j+1} \rho _j )\mathord{\buildrel{\lower3pt\hbox{$\scriptscriptstyle\leftrightarrow$}} 
		\over J} _\ell  (k_j \rho _j ) - J_\ell  (k_j \rho _j )\mathord{\buildrel{\lower3pt\hbox{$\scriptscriptstyle\leftrightarrow$}}
		\over H} _\ell ^{(1)} (k_{j+1} \rho _j )} \right]\mathord{\buildrel{\lower3pt\hbox{$\scriptscriptstyle\leftrightarrow$}}\over R} _{j,j+1}=\\
= H_\ell ^{(1)} (k_j \rho _j )\mathord{\buildrel{\lower3pt\hbox{$\scriptscriptstyle\leftrightarrow$}} 
	\over H} _\ell ^{(1)} (k_{j+1} \rho _j ) - H_\ell ^{(1)} (k_{j+1} \rho _j )\mathord{\buildrel{\lower3pt\hbox{$\scriptscriptstyle\leftrightarrow$}} 
	\over H} _\ell ^{(1)} (k_j \rho _j )
\end{multline}

\begin{equation}
\begin{array}{l}
\mathord{\buildrel{\lower3pt\hbox{$\scriptscriptstyle\leftrightarrow$}} 
	\over R} _{j,j+1}  =\\= \mathord{\buildrel{\lower3pt\hbox{$\scriptscriptstyle\leftrightarrow$}} 
	\over D} ^{-1}_{j,j+1} \left[ {H_\ell ^{(1)} (k_j \rho _j )\mathord{\buildrel{\lower3pt\hbox{$\scriptscriptstyle\leftrightarrow$}} 
		\over H} _\ell ^{(1)} (k_{j+1} \rho _j ) - H_\ell ^{(1)} (k_{j+1} \rho _j )\mathord{\buildrel{\lower3pt\hbox{$\scriptscriptstyle\leftrightarrow$}} 
		\over H} _\ell ^{(1)} (k_j \rho _j )} \right]\\
\mathord{\buildrel{\lower3pt\hbox{$\scriptscriptstyle\leftrightarrow$}} 
	\over T} _{j,j+1}  = \mathord{\buildrel{\lower3pt\hbox{$\scriptscriptstyle\leftrightarrow$}} 
	\over D} ^{ - 1}_{j,j+1} \left[ {\mathord{\buildrel{\lower3pt\hbox{$\scriptscriptstyle\leftrightarrow$}} 
		\over J} _\ell  (k_j \rho _j )H_\ell ^{(1)} (k_j \rho _j ) - J_\ell  (k_j \rho _j )\mathord{\buildrel{\lower3pt\hbox{$\scriptscriptstyle\leftrightarrow$}} 
		\over H} _\ell ^{(1)} (k_j \rho _j )} \right]\label{eq:rif}
\end{array}
\end{equation}
where
\begin{equation}
\mathord{\buildrel{\lower3pt\hbox{$\scriptscriptstyle\leftrightarrow$}} 
	\over D}_{j,j+1}  = H_\ell ^{(1)} (k_{j+1} \rho _j )\mathord{\buildrel{\lower3pt\hbox{$\scriptscriptstyle\leftrightarrow$}} 
	\over J} _\ell  (k_j \rho _j ) - J_\ell  (k_j \rho _j )\mathord{\buildrel{\lower3pt\hbox{$\scriptscriptstyle\leftrightarrow$}} 
	\over H} _\ell ^{(1)} (k_{j+1} \rho _j )
\end{equation}
\\
Analogously one can find the reflection and transmission matrices of a standing wave incident from the medium j+1 on the boundary $\rho_j$ as:
\begin{equation}
\begin{array}{l}
\mathord{\buildrel{\lower3pt\hbox{$\scriptscriptstyle\leftrightarrow$}} 
	\over R} _{j+1,j}  = \mathord{\buildrel{\lower3pt\hbox{$\scriptscriptstyle\leftrightarrow$}} 
	\over D} ^{-1}_{j,j+1} \\\left[ {J_\ell ^{(1)} (k_j \rho _j )\mathord{\buildrel{\lower3pt\hbox{$\scriptscriptstyle\leftrightarrow$}} 
		\over J} _\ell ^{(1)} (k_{j+1} \rho _j ) - J_\ell ^{(1)} (k_{j+1} \rho _j )\mathord{\buildrel{\lower3pt\hbox{$\scriptscriptstyle\leftrightarrow$}} 
		\over J} _\ell ^{(1)} (k_j \rho _j )} \right]\\
\mathord{\buildrel{\lower3pt\hbox{$\scriptscriptstyle\leftrightarrow$}} 
	\over T} _{j+1,j}  = \mathord{\buildrel{\lower3pt\hbox{$\scriptscriptstyle\leftrightarrow$}} 
	\over D} ^{ - 1}_{j,j+1} \\\left[ {\mathord{\buildrel{\lower3pt\hbox{$\scriptscriptstyle\leftrightarrow$}} 
		\over J} _\ell  (k_{j+1} \rho _j )H_\ell ^{(1)} (k_{j+1} \rho _j ) - J_\ell  (k_{j+1} \rho _j )\mathord{\buildrel{\lower3pt\hbox{$\scriptscriptstyle\leftrightarrow$}} 
		\over H} _\ell ^{(1)} (k_{j+1} \rho _j )} \right]\label{eq:rif2}
\end{array}
\end{equation}
\\
The function identities:
\begin{equation}
B_\ell ^{'} (x)=\pm B_{\ell \mp 1} (x) \mp \frac{\ell}{x} B_\ell (x)
\end{equation}
can be used to calculate the Wronskian of the Hankel functions, $H_\ell ^{(1)} (x)J_\ell ^{'} (x)-J_\ell (x)H_\ell ^{(1)'} (x)=-\frac{2i}{\pi x}$, and simplify the expression for $\mathord{\buildrel{\lower3pt\hbox{$\scriptscriptstyle\leftrightarrow$}} 
	\over T} _{j,j+1}$.
\\
Moreover the recurrence relations:
\begin{equation}
\begin{array}{l}
2B_\ell ^{'} (x)=B_{\ell-1} (x)-B_{\ell+1} (x)\\
2\frac{\ell}{x}B_\ell (x)=B_{\ell-1} (x)+B_{\ell+1} (x)
\end{array}
\end{equation}
allow to rewrite eq.~\ref{eq:eqb} in the form:
\begin{equation}
\mathord{\buildrel{\lower3pt\hbox{$\scriptscriptstyle\leftrightarrow$}} 
	\over B} _\ell  (k_j \rho ) =\frac{i}{2}\left[ B_{\ell-1} (k_j \rho)\mathord{\buildrel{\lower3pt\hbox{$\scriptscriptstyle\leftrightarrow$}} 
	\over M}+B_{\ell+1} (k_j \rho)\mathord{\buildrel{\lower3pt\hbox{$\scriptscriptstyle\leftrightarrow$}} 
	\over M}^{-1} \right]\label{eq:eqb2}
\end{equation}
where
$\mathord{\buildrel{\lower3pt\hbox{$\scriptscriptstyle\leftrightarrow$}}\over M}=\frac{1}{{k_j}}\left[ {\begin{array}{*{20}c}
	{\omega \varepsilon _j} & { i\beta  }\\
	{ i\beta } & { - \omega \mu _j }  \\
	\end{array}} \right]$
is a matrix with determinant 1, independent on $\ell$ and $\rho$.

\section{\label{sec:edge}Guided waves and edge modes}

Guided modes, defined as the nontrivial solutions that exist without the need for an external excitation, can be found requiring that the reflection matrix have an infinite determinant.
\\
For a two layer fiber they are found by imposing $\det (\mathord{\buildrel{\lower3pt\hbox{$\scriptscriptstyle\leftrightarrow$}}	\over D_{1,2}} ) = 0$. This follows from eq.~\ref{eq:rif} and for each $\ell$ value there will be m modes due to the periodic nature of the Bessel functions.
\\
The corresponding eigenfunctions are:
\begin{equation}
\begin{array}{l}
\left[ {H_\ell ^{(1)} (k_j \rho_j ) + J_\ell  (k_j \rho_j )R_{11}  - H_\ell ^{(1)} (k_{j+1} \rho_j )T_{11} } \right]e_{oz}=\\  =\left[ {H_\ell ^{(1)} (k_{j+1} \rho_j )T_{12}  - J_\ell  (k_j \rho_j )R_{12} } \right]h_{oz}  \\ 
\left[ {J_\ell  (k_j \rho_j )R_{21}  - H_\ell ^{(1)} (k_{j+1} \rho_j )T_{21} } \right]e_{oz}=\\  =\left[ {H_\ell ^{(1)} (k_{j+1} \rho_j )T_{22}  - H_\ell ^{(1)} (k_j \rho_j ) + J_\ell  (k_j \rho_j )R_{22} } \right]h_{oz}  \\ 
\end{array}
\end{equation}

\begin{equation}
\begin{array}{l}
1)\quad e_{oz}  = \left[ {H_\ell ^{(1)} (k_2 \rho _1 )T_{12}  - J_\ell  (k_1 \rho _1 )R_{12} } \right]\\ 
\quad \quad h_{oz}  = \left[ {H_\ell ^{(1)} (k_1 \rho _1 ) + J_\ell  (k_1 \rho _1 )R_{11}  - H_\ell ^{(1)} (k_2 \rho _1 )T_{11} } \right] \\ 
2)\quad e_{oz}  = \left[ {H_\ell ^{(1)} (k_2 \rho _1 )T_{22}  - H_\ell ^{(1)} (k_1 \rho _1 ) + J_\ell  (k_1 \rho _1 )R_{22} } \right]\\
\quad \quad h_{oz}  = \left[ {J_\ell  (k_1 \rho _1 )R_{21}  - H_\ell ^{(1)} (k_2 \rho _1 )T_{21} } \right] \\ 
\end{array}
\end{equation}
When $\ell  = 0$, modes are decoupled and can be classified as either transverse electric, TE (2) with $h_z=0$, or transvers magnetic, TM (1) with $e_z=0$.

For any mode a cutoff frequency is defined as the minimum frequency to have a positive $\beta ^2  = \frac{{\omega ^2 }}{{c^2 }}\varepsilon _j  - k_j^2 $  value i.e. a propagating wave.

\section{\label{sec:asym} Asymptotic analysis of the cladding}

The design of the topological fiber can be done through an asymptotic approach~\cite{Xu} that allows to analyze both guided and edge modes.
\\
If we use the asymptotic expressions for $k\rho\rightarrow \infty$ for the Bessel functions to describe the cladding, we can write the fields of eq.~\ref{eq:e2}) in the form:
\begin{equation}
\begin{array}{l}
\left[ {\begin{array}{*{20}c}
	{E_z^1 }  \\
	{H_z^1 }  \\
	\end{array}} \right] = J_\ell  (k_1 \rho )\vec a_1 \quad \quad \quad for \quad 0<\rho<\rho_1 \\
\\
\left[ {\begin{array}{*{20}c}
	{E_z^j }  \\
	{H_z^j }  \\
	\end{array}} \right] =\frac{1}{\sqrt{k_j\rho}}\left[e^{ik_j(\rho-\rho_{j-1})}\vec a_j+e^{-ik_j(\rho-\rho_{j-1})}\vec b_j\right]\quad \rho>\rho_1
\label{eq:asym2}
\end{array}
\end{equation}
\\
In particular in this limit we can write:
$\left[
{E_z^j }  ,	{H_z^j }  ,	{E_\theta^j } ,	{H_\theta^j }  \right]^T =
{\mathord{\buildrel{\lower3pt\hbox{$\scriptscriptstyle\leftrightarrow$}} 
		\over M_j } }\left[ 
{\vec a_j }  ,
{\vec b_j }  \right]^T$
where, for $j\ne 1$:
\begin{equation}
\begin{array}{l}
{\mathord{\buildrel{\lower3pt\hbox{$\scriptscriptstyle\leftrightarrow$}} 
		\over M_j } }(\rho) =\left[ {\begin{array}{*{20}c}
	{\mathord{\buildrel{\lower3pt\hbox{$\scriptscriptstyle\leftrightarrow$}} 
			\over A_{\ell j} } }&  	{\mathord{\buildrel{\lower3pt\hbox{$\scriptscriptstyle\leftrightarrow$}} 
			\over B_{\ell j} } }  \\
	{\mathord{\buildrel{\lower3pt\hbox{$\scriptscriptstyle\leftrightarrow$}} 
			\over C_{\ell j} } } & 	{\mathord{\buildrel{\lower3pt\hbox{$\scriptscriptstyle\leftrightarrow$}} 
			\over D_{\ell j} } }  \\
	\end{array}} \right]
\label{eq:asym3}
\end{array}
\end{equation}
with:
\begin{equation}
\begin{array}{l}
\mathord{\buildrel{\lower3pt\hbox{$\scriptscriptstyle\leftrightarrow$}} 
	\over A}_{ j}(\rho) =\frac{1}{\sqrt{k_j\rho}}e^{ik_j(\rho-\rho_{j-1})}\mathord{\buildrel{\lower3pt\hbox{$\scriptscriptstyle\leftrightarrow$}} 
	\over I} \\
\mathord{\buildrel{\lower3pt\hbox{$\scriptscriptstyle\leftrightarrow$}} 
	\over B}_{ j}(\rho)  =\frac{1}{\sqrt{k_j\rho}}e^{-ik_j(\rho-\rho_{j-1})}\mathord{\buildrel{\lower3pt\hbox{$\scriptscriptstyle\leftrightarrow$}} 
	\over I} \\
\mathord{\buildrel{\lower3pt\hbox{$\scriptscriptstyle\leftrightarrow$}} 
	\over C}_{ j}(\rho) =\frac{e^{ik_j(\rho-\rho_{j-1})}}{2k_j^2\rho \sqrt{k_j \rho }}
\left[ {\begin{array}{*{20}c}
	{ 2\ell \beta  } & {i\omega \mu (2ik_j \rho-1) } \\
	{-i\omega \epsilon_j (2ik_j \rho-1)}  & { 2\ell \beta }  \\
	\end{array}} \right]\\
\mathord{\buildrel{\lower3pt\hbox{$\scriptscriptstyle\leftrightarrow$}} 
	\over D}_{ j}(\rho) =\frac{e^{-ik_j(\rho-\rho_{j-1})}}{2k_j^2\rho \sqrt{k_j \rho }}
\left[ {\begin{array}{*{20}c}
	{ 2\ell \beta  } & {-i\omega \mu (2ik_j \rho+1) } \\
	{i\omega \epsilon_j (2ik_j \rho+1)}  & { 2\ell \beta }  \\
	\end{array}} \right]
\end{array}
\end{equation}
The boundary conditions at the interface between two adjacent layers (j,j+1) at $\rho=\rho_j$ then give:
\begin{equation}
\begin{array}{l}
\left[ {\begin{array}{*{20}c}
	{\vec a_{j+1} }  \\
	{\vec b_{j+1} }  \\
	\end{array}} \right]={\mathord{\buildrel{\lower3pt\hbox{$\scriptscriptstyle\leftrightarrow$}} 
		\over M_{{j+1}} ^{-1} } }(\rho_j){\mathord{\buildrel{\lower3pt\hbox{$\scriptscriptstyle\leftrightarrow$}} 
		\over M_{j} } }(\rho_j)\left[ {\begin{array}{*{20}c}
	{\vec a_j }  \\
	{\vec b_j }  \\
	\end{array}} \right]\equiv {\mathord{\buildrel{\lower3pt\hbox{$\scriptscriptstyle\leftrightarrow$}} 
		\over F } }(\rho_j)\left[ {\begin{array}{*{20}c}
	{\vec a_j }  \\
	{\vec b_j }  \\
	\end{array}} \right]
\label{eq:tjj}
\end{array}
\end{equation}
where, given:
\begin{equation}
\begin{array}{l}
{\mathord{\buildrel{\lower3pt\hbox{$\scriptscriptstyle\leftrightarrow$}} 
		\over \Delta_j } }(\rho_j) =\frac{2\omega}{k_{j+1}\sqrt{k_{j+1}\rho_j}}\left[ {\begin{array}{*{20}c}
	0&  	\mu  \\
	-\epsilon_{j+1} & 	0  \\
	\end{array}} \right]
\label{eq:asym3}
\end{array}
\end{equation}
and
${\mathord{\buildrel{\lower3pt\hbox{$\scriptscriptstyle\leftrightarrow$}} 
		\over X_j } }={\mathord{\buildrel{\lower3pt\hbox{$\scriptscriptstyle\leftrightarrow$}} 
		\over \Delta_j^{-1} } }{\mathord{\buildrel{\lower3pt\hbox{$\scriptscriptstyle\leftrightarrow$}} 
		\over C_{j+1} } }{\mathord{\buildrel{\lower3pt\hbox{$\scriptscriptstyle\leftrightarrow$}} 
		\over A_{j+1}^{-1} } }$
we have:
\begin{equation}
\begin{array}{l}
{\mathord{\buildrel{\lower3pt\hbox{$\scriptscriptstyle\leftrightarrow$}} 
		\over F } }(\rho_j)=\\
\left[ {\begin{array}{*{20}c}
	({\mathord{\buildrel{\lower3pt\hbox{$\scriptscriptstyle\leftrightarrow$}} 
			\over A_{j+1}^{-1} } }+{\mathord{\buildrel{\lower3pt\hbox{$\scriptscriptstyle\leftrightarrow$}} 
			\over X_j } }){\mathord{\buildrel{\lower3pt\hbox{$\scriptscriptstyle\leftrightarrow$}} 
			\over A_{j} } }-{\mathord{\buildrel{\lower3pt\hbox{$\scriptscriptstyle\leftrightarrow$}} 
			\over \Delta_{j}^{-1} } }{\mathord{\buildrel{\lower3pt\hbox{$\scriptscriptstyle\leftrightarrow$}} 
			\over C_{j} } }&  	({\mathord{\buildrel{\lower3pt\hbox{$\scriptscriptstyle\leftrightarrow$}} 
			\over A_{j+1}^{-1} } }+{\mathord{\buildrel{\lower3pt\hbox{$\scriptscriptstyle\leftrightarrow$}} 
			\over X_j } }){\mathord{\buildrel{\lower3pt\hbox{$\scriptscriptstyle\leftrightarrow$}} 
			\over B_{j} } }-{\mathord{\buildrel{\lower3pt\hbox{$\scriptscriptstyle\leftrightarrow$}} 
			\over \Delta_{j}^{-1} } }{\mathord{\buildrel{\lower3pt\hbox{$\scriptscriptstyle\leftrightarrow$}} 
			\over D_{j} } }  \\
	{\mathord{\buildrel{\lower3pt\hbox{$\scriptscriptstyle\leftrightarrow$}} 
			\over \Delta_{j}^{-1} } }({\mathord{\buildrel{\lower3pt\hbox{$\scriptscriptstyle\leftrightarrow$}} 
			\over C_{j} } } -{\mathord{\buildrel{\lower3pt\hbox{$\scriptscriptstyle\leftrightarrow$}} 
			\over C_{j+1} } }{\mathord{\buildrel{\lower3pt\hbox{$\scriptscriptstyle\leftrightarrow$}} 
			\over A_{j+1}^{-1} } }{\mathord{\buildrel{\lower3pt\hbox{$\scriptscriptstyle\leftrightarrow$}} 
			\over A_{j} } })& 	{\mathord{\buildrel{\lower3pt\hbox{$\scriptscriptstyle\leftrightarrow$}} 
			\over \Delta_{j}^{-1} } }({\mathord{\buildrel{\lower3pt\hbox{$\scriptscriptstyle\leftrightarrow$}} 
			\over D_{j} } }-{\mathord{\buildrel{\lower3pt\hbox{$\scriptscriptstyle\leftrightarrow$}} 
			\over C_{j+1} } }{\mathord{\buildrel{\lower3pt\hbox{$\scriptscriptstyle\leftrightarrow$}} 
			\over A_{j+1}^{-1} } }{\mathord{\buildrel{\lower3pt\hbox{$\scriptscriptstyle\leftrightarrow$}} 
			\over B_{j} } } ) \\
	\end{array}} \right]
\label{eq:tjj1}
\end{array}
\end{equation}
${\mathord{\buildrel{\lower3pt\hbox{$\scriptscriptstyle\leftrightarrow$}} \over F } }(\rho_j)$ is a block matrix that can be decomposed in a sum of an asymptotic contribution ${\mathord{\buildrel{\lower3pt\hbox{$\scriptscriptstyle\leftrightarrow$}} 
		\over F^\infty } }$ and a term for $\rho$ finite, ${\mathord{\buildrel{\lower3pt\hbox{$\scriptscriptstyle\leftrightarrow$}} 
		\over F^< } }$,
$
{\mathord{\buildrel{\lower3pt\hbox{$\scriptscriptstyle\leftrightarrow$}} 
		\over F } }(\rho_j)=
\left[ {\begin{array}{*{20}c}
	{\mathord{\buildrel{\lower3pt\hbox{$\scriptscriptstyle\leftrightarrow$}} 
			\over F_{11} } } &  {\mathord{\buildrel{\lower3pt\hbox{$\scriptscriptstyle\leftrightarrow$}} 
			\over F_{12} } } \\
	{\mathord{\buildrel{\lower3pt\hbox{$\scriptscriptstyle\leftrightarrow$}} 
			\over F_{21} } } &  {\mathord{\buildrel{\lower3pt\hbox{$\scriptscriptstyle\leftrightarrow$}} 
			\over F_{22} } }
	\end{array}} \right] \equiv {\mathord{\buildrel{\lower3pt\hbox{$\scriptscriptstyle\leftrightarrow$}} 
		\over F^< } } +	{\mathord{\buildrel{\lower3pt\hbox{$\scriptscriptstyle\leftrightarrow$}} 
		\over F^\infty } } 
$
where:
\begin{equation}
\begin{array}{l}
{\mathord{\buildrel{\lower3pt\hbox{$\scriptscriptstyle\leftrightarrow$}} 
		\over F^<_{11} } }(\rho_j)=\frac{e^{ik_j \ell_j}\sqrt{k_{j+1}}}{4\rho_jk_{j+1}\sqrt{k_j}}\left[ {\begin{array}{*{20}c}
	-i(1-\frac{k_{j+1}^2\epsilon_j}{k_j^2\epsilon_{j+1}}) & \frac{-2\ell\beta}{\omega\epsilon_{j+1}}(1-\frac{k_{j+1}^2}{k_j^2})  \\
	\frac{2\ell\beta}{\omega\mu}(1-\frac{k_{j+1}^2}{k_j^2})  & -i(1-\frac{k_{j+1}^2}{k_j}^2) \\
	\end{array}} \right]\\
{\mathord{\buildrel{\lower3pt\hbox{$\scriptscriptstyle\leftrightarrow$}} 
		\over F^<_{12} } }(\rho_j)=e^{-2ik_j \ell_j}{\mathord{\buildrel{\lower3pt\hbox{$\scriptscriptstyle\leftrightarrow$}} 
		\over F^<_{11} } }\\
{\mathord{\buildrel{\lower3pt\hbox{$\scriptscriptstyle\leftrightarrow$}} 
		\over F^<_{21} } }(\rho_j)=-{\mathord{\buildrel{\lower3pt\hbox{$\scriptscriptstyle\leftrightarrow$}} 
		\over F^<_{11} } }\\
{\mathord{\buildrel{\lower3pt\hbox{$\scriptscriptstyle\leftrightarrow$}} 
		\over F^<_{22} } }(\rho_j)=-{\mathord{\buildrel{\lower3pt\hbox{$\scriptscriptstyle\leftrightarrow$}} 
		\over F^<_{12} } }\\
\label{eq:tfinit}
\end{array}
\end{equation}
and:
\begin{equation}
\begin{array}{l}
{\mathord{\buildrel{\lower3pt\hbox{$\scriptscriptstyle\leftrightarrow$}} 
		\over F^\infty_{11} } }(\rho_j)=\frac{\sqrt{k_{j+1}}}{2\sqrt{k_j}}e^{ik_j \ell_j}\left[ {\begin{array}{*{20}c}
	1+\frac{k_{j+1}\epsilon_j}{k_j\epsilon_{j+1}} & 0  \\
	0 & 1+\frac{k_{j+1}}{k_j}  \\
	\end{array}} \right]\\
{\mathord{\buildrel{\lower3pt\hbox{$\scriptscriptstyle\leftrightarrow$}} 
		\over F^\infty_{12} } }(\rho_j)=\frac{\sqrt{k_{j+1}}}{2\sqrt{k_j}}e^{-ik_j \ell_j}\left[ {\begin{array}{*{20}c}
	1-\frac{k_{j+1}\epsilon_j}{k_j\epsilon_{j+1}} & 0  \\
	0 & 1-\frac{k_{j+1}}{k_j}  \\
	\end{array}} \right]\\\\
{\mathord{\buildrel{\lower3pt\hbox{$\scriptscriptstyle\leftrightarrow$}} 
		\over F^\infty_{21} } }(\rho_j)=\frac{\sqrt{k_{j+1}}}{2\sqrt{k_j}}e^{ik_j \ell_j}\left[ {\begin{array}{*{20}c}
	1-\frac{k_{j+1}\epsilon_j}{k_j\epsilon_{j+1}} & 0  \\
	0 & 1-\frac{k_{j+1}}{k_j}  \\
	\end{array}} \right]\\
{\mathord{\buildrel{\lower3pt\hbox{$\scriptscriptstyle\leftrightarrow$}} 
		\over F^\infty_{22} } }(\rho_j)=\frac{\sqrt{k_{j+1}}}{2\sqrt{k_j}}e^{-ik_j \ell_j}\left[ {\begin{array}{*{20}c}
	1+\frac{k_{j+1}\epsilon_j}{k_j\epsilon_{j+1}} & 0  \\
	0 & 1+\frac{k_{j+1}}{k_j}  \\
	\end{array}} \right]\\
\label{eq:tinf}
\end{array}
\end{equation}
\\
This decomposition allows us to identify, in the cladding, a $\rho$ value, $\rho_n$, such that for $\rho>\rho_n$, we can retain in the ${\mathord{\buildrel{\lower3pt\hbox{$\scriptscriptstyle\leftrightarrow$}} \over F } }$ matrix only the ${\mathord{\buildrel{\lower3pt\hbox{$\scriptscriptstyle\leftrightarrow$}} \over F^\infty } }$ terms.
\begin{figure}[t]
	\includegraphics[width=1\columnwidth]{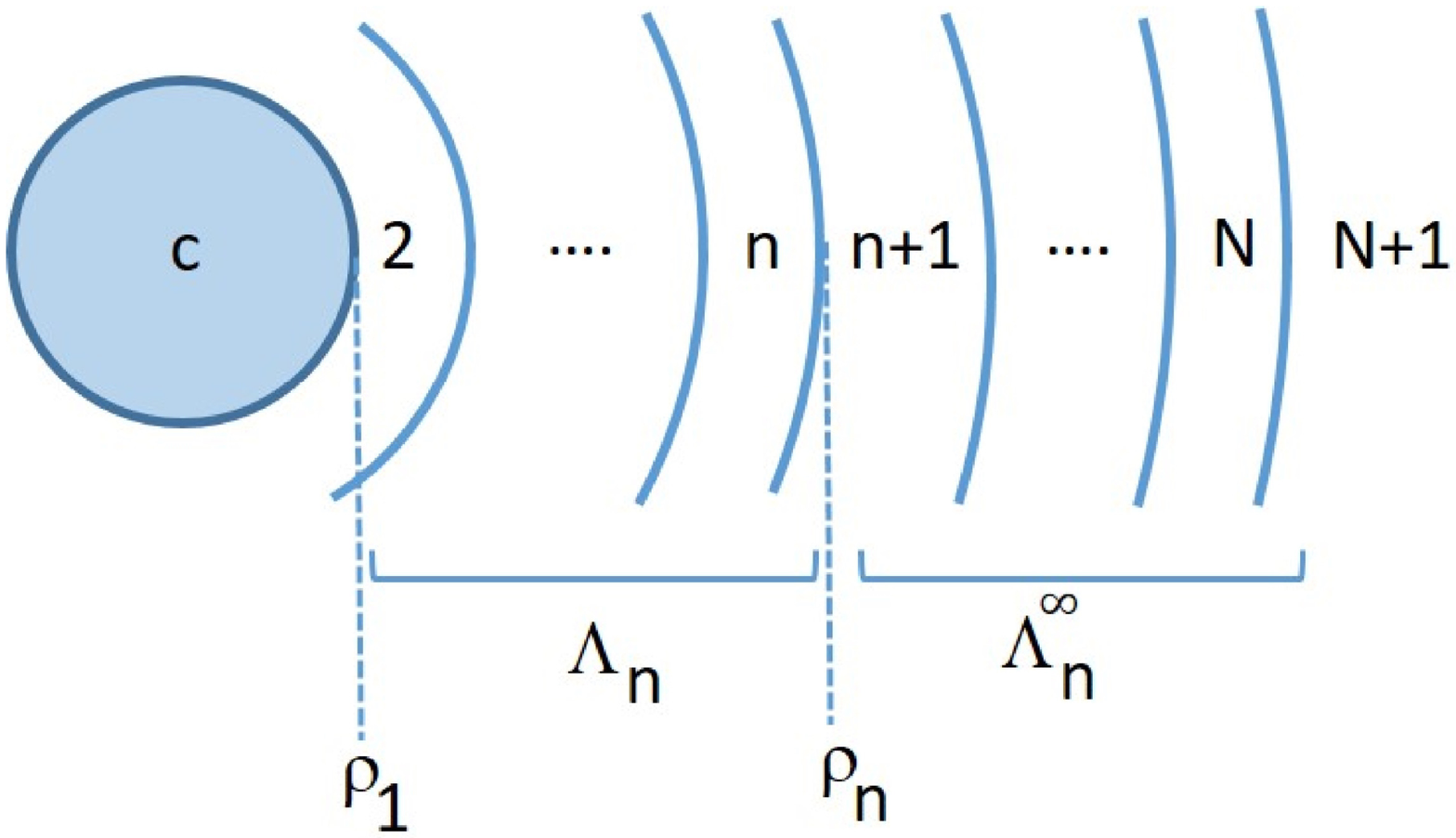}    	
	\caption{(Color online) Transfer matrix decomposition.}
\end{figure}
In this case the matrix in eq.~\ref{eq:tjj1} simplifies, no more depends on the $\rho_j$ coordinate and allows to separate the two polarizations TE:
\begin{equation}
\begin{array}{l}
{\mathord{\buildrel{\lower3pt\hbox{$\scriptscriptstyle\leftrightarrow$}} 
		\over F_{te} } }=\frac{\sqrt{k_{j+1}}}{2\sqrt{k_j}}
\left[ {\begin{array}{*{20}c}
	e^{ik_j \ell_j}(1+\frac{k_{j+1}}{k_j})  & e^{-ik_j \ell_j}(1-\frac{k_{j+1}}{k_j}) \\
	e^{ik_j \ell_j}(1-\frac{k_{j+1}}{k_j})  &  e^{-ik_j \ell_j}(1+\frac{k_{j+1}}{k_j}) \\
	\end{array}} \right]
\label{eq:tjjs1}
\end{array}
\end{equation}
\\
and TM:
\begin{equation}
\begin{array}{l}
{\mathord{\buildrel{\lower3pt\hbox{$\scriptscriptstyle\leftrightarrow$}} 
		\over F_{tm} } }=\\\frac{\sqrt{k_{j+1}}}{2\sqrt{k_j}}
\left[ {\begin{array}{*{20}c}
	e^{ik_j \ell_j}(1+\frac{k_{j+1}\epsilon_j}{k_j\epsilon_{j+1}})  &  e^{-ik_j \ell_j}(1-\frac{k_{j+1}}{k_j}\frac{\epsilon_j}{\epsilon_{j+1}}) \\
	e^{ik_j \ell_j}(1-\frac{k_{j+1}\epsilon_j}{k_j\epsilon_{j+1}})  &  e^{-ik_j \ell_j}(1+\frac{k_{j+1}}{k_j}\frac{\epsilon_j}{\epsilon_{j+1}}) \\
	\end{array}} \right]
\label{eq:tjjs2}
\end{array}
\end{equation}
\\
The same is true for the transfer matrix on the single period of the cladding, so that we can define the ${\mathord{\buildrel{\lower3pt\hbox{$\scriptscriptstyle\leftrightarrow$}}\over S_{te} } }$ and ${\mathord{\buildrel{\lower3pt\hbox{$\scriptscriptstyle\leftrightarrow$}}\over S_{tm} } }$ matrices and obtain the gaps of the cladding by the half-trace $x_{te/tm}=\frac{1}{2}Tr({\mathord{\buildrel{\lower3pt\hbox{$\scriptscriptstyle\leftrightarrow$}}\over S_{te/tm} }})$ of the transfer matrix.
\\
So, given the decomposition $
{\mathord{\buildrel{\lower3pt\hbox{$\scriptscriptstyle\leftrightarrow$}} 
		\over F } }(\rho_j)=
{\mathord{\buildrel{\lower3pt\hbox{$\scriptscriptstyle\leftrightarrow$}} 
		\over F^< } } +	{\mathord{\buildrel{\lower3pt\hbox{$\scriptscriptstyle\leftrightarrow$}} 
		\over F^\infty } }$
and a structure with N layers, we have:
\begin{equation}
\begin{array}{l}
\left[ {\begin{array}{*{20}c}
	{\vec a_{N+1} }  \\
	{\vec b_{N+1} }  \\
	\end{array}} \right]=\prod_{i=0}^{N-n-1}{\mathord{\buildrel{\lower3pt\hbox{$\scriptscriptstyle\leftrightarrow$}} 
		\over F^\infty } }(\rho_{N-i})\left[ {\begin{array}{*{20}c}
	{\vec a_{n+1} }  \\
	{\vec b_{n+1} }  \\
	\end{array}} \right]\equiv{\mathord{\buildrel{\lower3pt\hbox{$\scriptscriptstyle\leftrightarrow$}} 
		\over \Lambda^\infty_n } }\left[ {\begin{array}{*{20}c}
	{\vec a_{n+1} }  \\
	{\vec b_{n+1} }  \\
	\end{array}} \right]\\
\left[ {\begin{array}{*{20}c}
	{\vec a_{n+1} }  \\
	{\vec b_{n+1} }  \\
	\end{array}} \right]=\prod_{i=0}^{n-2}{\mathord{\buildrel{\lower3pt\hbox{$\scriptscriptstyle\leftrightarrow$}} 
		\over F } }(\rho_{n-i})\left[ {\begin{array}{*{20}c}
	{\vec a_2 }  \\
	{\vec b_2 }  \\
	\end{array}} \right]\equiv{\mathord{\buildrel{\lower3pt\hbox{$\scriptscriptstyle\leftrightarrow$}} 
		\over \Lambda_n } }\left[ {\begin{array}{*{20}c}
	{\vec a_2 }  \\
	{\vec b_2 }  \\
	\end{array}} \right]
\label{eq:clad}
\end{array}
\end{equation}
Obviously, if the cladding region between n+1 and N is composed of p periods, ${\mathord{\buildrel{\lower3pt\hbox{$\scriptscriptstyle\leftrightarrow$}} 
		\over \Lambda^\infty_n } }={\mathord{\buildrel{\lower3pt\hbox{$\scriptscriptstyle\leftrightarrow$}} 
		\over S^p } }$.
\\
The last condition is at the core-cladding interface, $\rho=\rho_1$, between the exact solution in the core:
\begin{equation}
\begin{array}{l}
\left[ {\begin{array}{*{20}c}
	{E_z^1 }  \\
	{H_z^1 }  \\
	{E_\theta^1 }  \\
	{H_\theta^1 }  \\
	\end{array}} \right] = \left[ {\begin{array}{*{20}c}
	J_\ell  (k_1 \rho )&  	0  \\
	0 & 	J_\ell  (k_1 \rho ) \\
	{ - \ell \beta J_\ell  (k_1 \rho )/(k_1^2 \rho )} & { - i\omega \mu J_\ell' (k_1 \rho )/k_1}\\
	{i\omega \varepsilon _1 J_\ell'(k_1 \rho )/k_1}& { - \ell \beta J_\ell  (k_1 \rho )/(k_1^2 \rho ) }  \\
	\end{array}} \right]\vec a_1 
\label{eq:asy}
\end{array}
\end{equation}
and the asymptotic one in the cladding.
\\
It reads:
\begin{equation}
\begin{array}{l}
{\mathord{\buildrel{\lower3pt\hbox{$\scriptscriptstyle\leftrightarrow$}} 
		\over M_1 } }(\rho_1 )\vec a_1 ={\mathord{\buildrel{\lower3pt\hbox{$\scriptscriptstyle\leftrightarrow$}} 
		\over M_2 } }(\rho_1 )\left[ {\begin{array}{*{20}c}
	{\vec a_2 }  \\
	{\vec b_2 }  \\
	\end{array}} \right]
\label{eq:asyc}
\end{array}
\end{equation}
From eq.~\ref{eq:clad} we have:
\begin{equation}
\begin{array}{l}
\left[ {\begin{array}{*{20}c}
	{\vec a_{N+1} }  \\
	{\vec b_{N+1} }  \\
	\end{array}} \right]={\mathord{\buildrel{\lower3pt\hbox{$\scriptscriptstyle\leftrightarrow$}} 
		\over \Lambda^\infty_n } }{\mathord{\buildrel{\lower3pt\hbox{$\scriptscriptstyle\leftrightarrow$}} 
		\over \Lambda_n } }
{\mathord{\buildrel{\lower3pt\hbox{$\scriptscriptstyle\leftrightarrow$}} 
		\over M_2^{-1} } }{\mathord{\buildrel{\lower3pt\hbox{$\scriptscriptstyle\leftrightarrow$}} 
		\over M_1 } }\vec a_1
\label{eq:claa}
\end{array}
\end{equation}
where:
\begin{equation}
\begin{array}{l}
{\mathord{\buildrel{\lower3pt\hbox{$\scriptscriptstyle\leftrightarrow$}} 
		\over M_1 } }=\left[ {\begin{array}{*{20}c}
	J_\ell  (k_1 \rho_1 )&  	0  \\
	0 & 	J_\ell  (k_1 \rho_1 ) \\
	{ - \ell \beta J_\ell  (k_1 \rho_1 )/(k_1^2 \rho_1 )} & { - i\omega \mu _1 J_\ell' (k_1 \rho_1 )/k_1}\\
	{i\omega \varepsilon _1 J_\ell'(k_1 \rho_1 )/k_1}& { - \ell \beta J_\ell  (k_1 \rho_1 )/(k_1^2 \rho_1 ) }  \\
	\end{array}} \right]
\label{eq:asyl}
\end{array}
\end{equation}
If we consider that
$\vec b_j=
\tilde {\mathord{\buildrel{\lower3pt\hbox{$\scriptscriptstyle\leftrightarrow$}} 
		\over R}} _{j,j+1} \vec a_j$ and that $\tilde {\mathord{\buildrel{\lower3pt\hbox{$\scriptscriptstyle\leftrightarrow$}} 
		\over R}} _{N,N+1}=0$ then eq.~\ref{eq:asyc} gives four equations in the four variables $(e_{1z},h_{1z},e_{2z},h_{2z})$ while eq.~\ref{eq:claa} allows to determine the propagation constant $\beta$ and the field distribution of guided modes.
\\
For modes with $\ell=0$ we notice that also the component for $\rho$ finite, ${\mathord{\buildrel{\lower3pt\hbox{$\scriptscriptstyle\leftrightarrow$}}\over F^< } }$, of the transfer matrix is block diagonalized into 2x2 matrices and, as expected, the polarizations are not coupled.
\\
We can rewrite eq.~\ref{eq:claa} in the form:
\begin{equation}
\begin{array}{l}
\left[ {\begin{array}{*{20}c}
	{\vec a_{N+1} }  \\
	{\vec b_{N+1} }  \\
	\end{array}} \right]={\mathord{\buildrel{\lower3pt\hbox{$\scriptscriptstyle\leftrightarrow$}} 
		\over \Xi } }\vec a_1
\label{eq:cla}
\end{array}
\end{equation}
where ${\mathord{\buildrel{\lower3pt\hbox{$\scriptscriptstyle\leftrightarrow$}} 
		\over \Xi } }$ is a 4x2 matrix, so that the condition for the TE guided modes is $\Xi_{31}=0$ and for the TM modes $\Xi_{42}=0$.
\\
The conditions $a_{e(N+1)} =\Xi_{11}a_{e1}$ and $a_{h(N+1)} =\Xi_{41}a_{h1}$ allow to determine the fields in the whole structure. Normalizing the fields to the core amplitudes $a_{e1}=1$ or $a_{h1}=1$ we have $a_{e(N+1)} =\Xi_{11}$ and $a_{h(N+1)} =\Xi_{41}$.


\begin{thebibliography}{99}

\bibitem{Raghu}
S. Raghu,  F. D. M. Haldane, "Analogs of quantum-hall-effect edge states in photonic crystals" \ Phys. \ Rev. \ A. \textbf{78}, 033834 (2008).

\bibitem{Haldane}
F. D. M. Haldane, S. Raghu, "Possible realization of directional optical waveguides in photonic crystals with broken time-reversal symmetry" \ Phys. \ 
Rev. \ Lett. \textbf{100}, 013904 (2008).

\bibitem{Wang}
Z. Wang et al., "Reflection-Free One-Way Edge Modes in a Gyromagnetic Photonic Crystal" \ Phys. \ 
Rev. \ Lett., \textbf{100}, 013905 (2008)

\bibitem{Wang2}
Z. Wang, et al., "Observation of unidirectional backscattering-immune topological electromagnetic states" Nature, \textbf{461}, 772 (2009)

\bibitem{Hafezi1}
M. Hafezi et al., "Robust optical delay lines with topological protection" Nature Physics, \textbf{7}, 907 (2011)

\bibitem{Fang}
K. Fang, Z. Yu, and S. Fan, "Realizing effective magnetic field for photons by controlling the phase of
dynamic modulation" Nature Photonics, \textbf{6}, 782 (2012)

\bibitem{Khanikaev}
A. B. Khanikaev, et al., "Photonic topological insulators" Nature Materials, \textbf{12}, 233 (2012)

\bibitem{Skirlo}
S.A. Skirlo, L. Lu, M. Soljacic, "Multimode One-Way Waveguides of Large Chern Numbers"
\ Phys. \  Rev. \ Lett., \textbf{113}, 113904 (2014)

\bibitem{Hafezi2}
M. Hafezi, et al., "Imaging topological edge states in silicon photonics" Nature Photonics, \textbf{7}, 1001 (2013) 


\bibitem{Rechtsman}
M. C. Rechtsman, et al. "Photonic floquet topological insulators" Nature \textbf{496}, 196 (2013).

\bibitem{Longhi}
S. Longhi, "Topological Phase Transition in non-Hermitian Quasicrystals", \ Phys. \ 
Rev. \ Lett., \textbf{122}, 237601 (2019)

\bibitem{Zeng}
Qi-Bo Zeng, Yan-Bin Yang, Yong Xu, "Topological Phases in Non-Hermitian Aubry-Andre-Harper Models", arXiv:1901.08060 

\bibitem{Pilozzi}
L. Pilozzi, C. Conti, "Topological lasing in resonant photonic structures" \ Phys. \ Rev. \ B \textbf{93} 195317 (2016)

\bibitem{Amo}
P. St-Jean, V. Goblot, E. Galopin, A. Lemaître, T. Ozawa, L. Le Gratiet, I. Sagnes, J. Bloch, A. Amo "Lasing in topological edge states of a one-dimensional lattice"
Nature Photonics \textbf{11}, 651 (2017) 

\bibitem{Bahari}
B. Bahari, A. Ndao, F. Vallini, A. El Amili, Y. Fainman, B. Kante, "Nonreciprocal lasing in topological cavities of arbitrary geometries" Science  \textbf{358}, 636 (2017)

\bibitem{Hararil}

G. Harari, M. A. Bandres, Y. Lumer, M. C. Rechtsman, Y. D. Chong, M. Khajavikhan, D. N. Christodoulides, M. Segev, "Topological insulator laser: Theory"
Science  \textbf{359} (2018)

\bibitem{Pilozzi2017}
L. Pilozzi, C. Conti, "Topological cascade laser for frequency comb generation in PT-symmetric structures"
\ Opt. \ Lett.   \textbf{42}, 5174 (2017)

\bibitem{Kruk2019}
S. Kruk et al.,"Nonlinear light generation in topological nanostructures" Nature Nanotechnology \textbf{14}, 126 (2019)

\bibitem{Koshelev2019}
K. Koshelev, G. Favraud, A. Bogdanov, Y. Kivshar, A. Fratalocchi, "Nonradiating photonics with resonant dielectric nanostructures", Nanophotonics \textbf{8}, 725 (2019)

\bibitem{Mittal}
S. Mittal, V. Vikram Orre, and M. Hafezi "Topologically robust transport of entangled photons
in a 2d photonic system" Opt. Express \textbf{24},15631 (2016)

\bibitem{Rechtsman2}
M. C. Rechtsman, Y. Lumer, Y. Plotnik, A.
Perez-Leija, A. Szameit, and M. Segev, "Topological protection of photonic path entanglement"
Optica \textbf{3}, 925 (2016).

\bibitem{Mittal3}
S. Mittal, E.A. Goldschmidt, M. Hafezi, "A topological source of quantum light"
Nature \textbf{561}, 502 (2018)

\bibitem{Pilozzi2018}
L. Pilozzi, F.A. Farrelly, G. Marcucci, C. Conti, "Machine learning inverse problem for topological photonics", Communication Physics \textbf{1}, 57 (2018)

\bibitem{Long2019}
Y. Long, J. Ren, Y. Li, and H. Chen, "Inverse design of photonic topological state via machine learning", \ Appl. \ Phys. \ Lett. \textbf{114},181105 (2019)

\bibitem{Thouless}
D. J. Thouless, M. Kohmoto, M. P. Nightingale, M. den Nijs, "Quantized Hall Conductance in a Two-Dimensional Periodic Potential". \ Phys. \ Rev. \ Lett. \textbf{49}, 405 (1982)

\bibitem{Hatsugai}
Y. Hatsugai "Edge states in the integer quantum Hall effect and the Riemann surface of the Bloch function"
Phys. Rev. B \textbf{48}, 11851 (1993)

\bibitem{Ozawa}
T. Ozawa, H. M. Price, N. Goldman, O. Zilberberg, and I. Carusotto, "Synthetic dimensions in integrated photonics: from optical isolation to four-dimensional quantum Hall physics", \ Phys. \ Rev. A \textbf{93}, 043827 (2016

\bibitem{Yuan}
L. Yuan, Q. Lin, M. Xiao, and S. Fan, "Synthetic dimension in photonics"
Optica \textbf{5}, 1396, (2018) 	

\bibitem{Lustig}	
E. Lustig, S. Weimann, Y. Plotnik, Y. Lumer, Miguel A. Bandres, A. Szameit, M. Segev
"Photonic topological insulator in synthetic dimensions" Nature \textbf{567}, 356 (2019) 

\bibitem{Harper}
P. G. Harper, "The General Motion of Conduction Electrons in a Uniform Magnetic Field, with Application to the Diamagnetism of Metals", \ Proc. \ Phys. \ Soc., \ London, Sect. A  \textbf{68}, 874 (1955)

\bibitem{Aubry}
S. Aubry and G. Andre, "Analicity breaking and Anderson localization in incommensurate lattices" \ Ann. \ Isr. \ Phys. \ Soc. \textbf{3}, 133 (1980).

\bibitem{Kraus}
Y. E. Kraus, Y. Lahini, Z. Ringel, M. Verbin, O. Zilberberg "Topological States and Adiabatic Pumping in Quasicrystals",  \ Phys. \ Rev. \ Lett. \textbf{109},106402 (2012)

\bibitem{Ganeshan}
S. Ganeshan, K. Sun, S. Das Sarma, "Topological Zero-Energy Modes in Gapless Commensurate Aubry-Andre-Harper Models", \ Phys. \ Rev. \ Lett. \textbf{110},180403 (2013)

\bibitem{Hofstadter}
D. R. Hofstadter, "Energy levels and wave functions of Bloch electrons in rational and irrational magnetic fields",  \ Phys. \ Rev. \ B \textbf{14}, 2239 (1976)

\bibitem{Posha}
A. V. Poshakinskiy, A. N. Poddubny, L. Pilozzi, E. L. Ivchenko, "Radiative topological states in resonant photonic crystals", \ Phys. \ Rev., \ Lett.,  \textbf{112}, 107403 (2014)

\bibitem{Yariv}
P. Yeh, A. Yariv, and E. Marom, "Theory of Bragg fiber" \ J. \ Opt. \ Soc. \ Am. \textbf{68}, 1196 (1978).

\bibitem{Johnson}
S.~G.~Johnson, M.~Ibanescu, M.~Skorobogatiy, O.~Weisberg, T.~D.~Engeness, M.~Soljacic, S.~A.~Jacobs, J.~D.~Joannopoulos, and Y.~Fink, "Low-loss asymptotically single-mode propagation in large-core OmniGuide fibers"
Opt.~Express \textbf{9}, 748 (2001).

\bibitem{Knight}
J.C. Knight, J. Broeng, T.A. Birks, "Photonic band gap guidance in optical fibers"
Science \textbf{282}, (1998)

\bibitem{Russel}
P. Russell, "Photonic crystal fibers" Science, \textbf{299}, (2003)

\bibitem{Cregan}
R.F. Cregan, B.J. Mangan, J.C. Knight, T.A. Birks "Single-mode photonic band gap guidance of light in air"
Science, \textbf{285} (1999)

\bibitem{Wang2006}  
X. Wang, Z. Chen, and J. Yang
"Guiding light in optically induced ring lattices with a low-refractive-index core"
\ Opt. \ Lett. \textbf{31}, 1887 (2006)

\bibitem{Lu2018}
L. Lu, H. Gao, Z. Wang
"Topological one-way fiber of second Chern number"
Nature Communications \textbf{9}, 5384 (2018)


\bibitem{Wong2012}
G. Wong et al., "Excitation of Orbital Angular Momentum Resonances in Helically Twisted Photonic Crystal Fiber",  Science \textbf{337}, 6093 (2012)

\bibitem{butsch2012}
A. Butsch, C. Conti, F. Biancalana, and P.St.J. Russel, "Optomechanical Self-Channeling of Light in a Suspended Planar Dual-Nanoweb Waveguide", \ Phys. \ Rev. \ Lett. \textbf{108}, 093903 (2012)

\bibitem{garbos2011}
M.~K. Garbos, T.~G. Euser, O.~A. Schmidt, S. Unterkofler, P.~S. Russell, "Doppler velocimetry on microparticles trapped and propelled by laser light in liquid-filled photonic crystal fiber", Opt. Lett. \textbf{11}, 2020 (2011)

\bibitem{Scheuer}
J. Scheuer and A. Yariv, "Annular Bragg defect mode resonators" \ J. \ Opt. \ Soc. \ Amer. \ B, \textbf{20}, 2285 (2003).

\bibitem{Xu}
Y. Xu, R.K. Lee and A. Yariv, "Asymptotic analysis of Bragg fibers," \ Opt. \ Lett. \textbf{25}, 1756 (2000).

\bibitem{Xu2}
Y. Xu, G. X. Ouyang, R. K. Lee, A. Yariv, "Asymptotic Matrix Theory of Bragg Fibers", Journal of Lightwave technology,  \textbf{20}, 428 (2002)

\bibitem{Chew}
W. C. Chew, "Waves and Fields in Inhomogeneous Media" New York: Van Nostrand Reinhold, 1990.

\bibitem{Posha2}
A. V. Poshakinskiy, A. N. Poddubny, and M. Hafezi, "Phase spectroscopy of topological invariants in photonic crystals", \ Phys. \ Rev. A \textbf{91}, 043830 (2015). 

\bibitem{Xu1}
Y. Xu, W. Liang, A. Yariv, J. G. Fleming and Shawn-Yu Lin "High-quality-factor Bragg onion resonators with
omnidirectional reflector cladding" \ Opt. \ Lett. \textbf{28}, 2144 (2003)

\bibitem{Huang}
K. C. Huang, E. Lidorikis, X. Jiang, J. D. Joannopoulos, and K. A. Nelson, "Nature of lossy Bloch states in polaritonic photonic crystals", \ Phys. \ Rev. B \textbf{69}, 195111 (2004).

\bibitem{Hasan}
M. Z. Hasan and C. L. Kane, "Colloquium: Topological insulators", \ Rev. \ Mod. \ Phys. \textbf{82}, 3045 (2010)
	
\end{thebibliography}
\end{document}